\documentclass[photonics,article,accept,moreauthors,10pt,a4paper]{mdpi} 

\firstpage{1} 
\articlenumber{60}
\doinum{10.3390/photonics3040060}
\pubvolume{3}
\pubyear{2016}
\copyrightyear{2016}
\history{Received: 31 October 2016; Accepted: 18 November 2016; Published: 23 November 2016}

\pdfoutput=1

\usepackage{amsmath,amsfonts}

\usepackage{booktabs} 
\usepackage{multirow}
\usepackage{soul}
\usepackage{microtype}

\usepackage{upgreek}

\newcommand{\Eth}{\vec{\mathcal{E}}}
\newcommand{\Dth}{\vec{\mathcal{D}}}
\newcommand{\Hth}{\vec{\mathcal{H}}}
\newcommand{\Bth}{\vec{\mathcal{B}}}

\newcommand{\chith}{\mathfrak{X}}
\newcommand{\chielth}{\mathfrak{X}_\text{e}}
\newcommand{\chimath}{\mathfrak{X}_\text{m}}
\newcommand{\Sigmth}{{\vec{\mathfrak{S}}}}

\newcommand{\Uelth}{\mathcal{U}_\text{e}}
\newcommand{\Umath}{\mathcal{U}_\text{m}}
\newcommand{\Uth}{\mathcal{U}}
\newcommand{\URth}{\mathcal{U}_\text{R}}
\newcommand{\ULth}{\mathcal{U}_\text{L}}

\newcommand{\E}{\vec{E}}
\newcommand{\D}{\vec{D}}
\renewcommand{\H}{\vec{H}}
\newcommand{\B}{\vec{B}}
\newcommand{\J}{\vec{J}}

\renewcommand{\S}{\vec{S}}
\newcommand{\UU}{U}

\newcommand{\chii}{\chi}
\newcommand{\Sigm}{\vec{\Sigma}}

\newcommand{\xx}{\vec{x}}

\newcommand{\epsi}{\varepsilon}
\newcommand{\mue}{\mu}

\newcommand{\nabl}{\vec{\nabla}}
\newcommand{\rot}[1]{\left( \nabl \times #1 \right)}

\newcommand{\imag}[1]{\operatorname{Im}\left(#1\right)}

\renewcommand{\vec}[1]{\boldsymbol{#1}}

\newcommand{\dt}{\partial_t}

\newcommand{\revChange}[1]{#1}
\newcommand{\revChangeT}[1]{#1}

\Title{Locally Enhanced and Tunable Optical Chirality in Helical Metamaterials}

\Author{Philipp Gutsche $^{1,}$*, Raquel M\"ausle $^{1}$ and Sven Burger $^{1,2}$}

\AuthorNames{Philipp Gutsche, Raquel M\"ausle and Sven Burger}

\address{%
$^{1}$ \quad Zuse Institute Berlin, Takustr. 7, 14195 Berlin, Germany; maeusle@zib.de (R.M.); burger@zib.de  (S.B.)\\
$^{2}$ \quad JCMwave GmbH, Bolivarallee 22, 14050 Berlin, Germany}

\corres{Correspondence: gutsche@zib.de; Tel.: +49-30-841-85-203}

\abstract{
We report on a numerical study of optical chirality. 
Intertwined gold helices illuminated with plane waves concentrate right and left circularly polarized electromagnetic field energy to sub-wavelength regions. 
These spots of enhanced chirality can be smoothly shifted in position and magnitude by varying illumination parameters, 
allowing for the control of light-matter interactions on a~nanometer scale. 
}


\begin{document}

\section{Introduction}
Helical metamaterials strongly impact the optical response of incident
circularly (CPL) and linearly polarized light. 
They serve as efficient circular polarizers \cite{gansel2009}
and are candidates for chiral near-field sources \cite{schaeferling2014}.
Both, advancement of fabrication techniques \cite{kaschke2016,deng2016,esposito2014} 
and design which employs fundamental physical properties \cite{kaschke2012}
have significantly increased the performance of these complex~structures.

Most experimental, numerical and theoretical studies focus on the far-field response of helical metamaterials.
Nevertheless, near-field interaction of light and chiral matter is expected to be enhanced in chiral near-fields. 
Recently, the quantity of optical chirality has been introduced quantifying this phenomenon \cite{tang2010}.
In the weak-coupling regime, the interplay of 
electromagnetic fields and chiral molecules, which are not superimposable with their mirror image, 
is directly proportional to this near-field measure \cite{schaeferling2016}.

Here, we use the concept of circularly polarized energy (CPE) which follows from the relation of optical chirality and
electromagnetic field energy for lossless isotropic media \cite{bliokh2011}.
We recapitulate the definitions of right and left CPE parts \cite{gutsche2016} and give a novel formal derivation
of conservation of optical chirality in arbitrary space.
The numerically investigated metamaterial consists of intertwined and tapered gold 3-helices. 
We observe subwavelength concentration of optical chirality density enhancement of more than a factor of four. 
Spatial control on a range of 1$\upmu$m is achieved while maintaining stable near-field intensities.

\section{Optical Chirality and Circularly Polarized Energy}

Analyzing light-matter interaction is mostly done with the help of electric and magnetic dipole moments of matter coupling
to electric and magnetic parts of light, respectively. Emerging interest in specifically chiral matter yielded
the introduction of the optical chirality density
$\chi = \epsi_0/2 \E \cdot \rot{\E} + 1/(2\mue_0) \B \cdot \rot{\B}$ \cite{tang2010}
with the electric field $\E$, the magnetic flux density $\B$,
vacuum permittivity $\epsi_0$ and permeability $\mue_0$.

This time-even pseudoscalar resembles symmetry properties of chiral shapes: it changes sign
under spatial inversion but not under time reversal. It
occurs together with the mixed electric-magnetic
dipole moment in the absorption rate of molecules and represents especially coupling of electromagnetic fields to chiral molecules.
Furthermore, it satisfies the conservation law $\partial_t \chi + \nabl \cdot \Sigm = 0$
with its flux $\Sigm = 1/2[\E \times \rot{\B} - \B \times \rot{\E}]$
in source-free, isotropic and homogeneous space.

Recently, this conservation law has been extended to arbitrary source- and material-distributions~\cite{gutsche2016,poulikakos2016}.
The occurring modified definitions of chiral quantities follow directly from Poynting's theorem.
Formally, the relation
$	f:
	\vec{a} \cdot \vec{b}
	\mapsto
	1/2 [\rot{\vec{a}} \cdot \vec{b}
	+
	\vec{a} \cdot \rot{\vec{b}}]
$
acting on a scalar product $\vec{a} \cdot \vec{b}$ transforms conventional energy to chirality conservation:
\begin{align}
	&f\left\{ \dt \UU + \nabl \cdot \S
		=  \J \cdot \E \right\}
	~\mapsto~ 
	\dt \chii + \nabl \cdot \Sigm
		= 1/2 \left[ \rot{\J} \cdot \E + \J \cdot \rot{\E} \right],
			\label{eq:ch}
\end{align}
where $\UU = U_e + U_m = 1/2 (\E \cdot \D + \H \cdot \B)$ is the field energy consisting
of an electric and magnetic part, respectively. The electric displacement field is $\D = \epsi \E$ and $\B = \mue \H$ with the magnetic field $\H$. $\J$ is the free current density and $\S= \E \times \H$ is the energy flux.

Here, we study monochromatic fields governed by Maxwell's equations \cite{jackson1998}.
We consider complex time-harmonic electric fields
$\Eth = \E \exp{(-i \omega t)}$ with angular frequency $\omega = 2 \pi \nu$. Complex-valued material parameters $\epsi$ and $\mue$ account for losses
and the time-averaged energy densities are denoted with $\Uth$.
The time-harmonic optical chirality in free space reads
$\chith = -\epsi_0 \omega / 2 \imag{\Eth^* \cdot \Bth}$ \cite{schaeferling2012}.
The~generalized continuity Equation \eqref{eq:ch} introduces an electric $\chielth$ and magnetic $\chimath$
chirality density
\begin{align}
	\begin{gathered}
	\chielth = 1/8 \left[ \Dth^* \cdot \rot{\Eth} + \Eth \cdot \rot{\Dth^*} \right],
	~~~
	\chimath = 1/8 \left[ \Hth^* \cdot \rot{\Bth} + \Bth \cdot \rot{\Hth^*} \right] \\
	\Sigmth = 1/4 \left[ \Eth \times \rot{\Hth^*} - \Hth^* \times \rot{\Eth} \right]
	\end{gathered}
	\label{eq:ch}
\end{align}
with $\chith = \chielth + \chimath$ and the optical chirality flux $\Sigmth$.

The optical chirality is proportional to the difference of CPE
in isotropic lossless materials with refractive index $n$ and speed of light $c=c_0/n$ with vacuum value $c_0$ \cite{bliokh2011}.
This gives rise to a basis change for the total energy $\Uth$ from electric and magnetic parts to right $\URth$ and
left $\ULth$ circularly polarized parts:
\begin{align}
	\chith(\xx) &= \frac{\omega n}{c_0} \left[ \ULth(\xx) - \URth(\xx) \right],
	\label{eq:chDiff}
	\\
	\Uth(\xx) &= \ULth(\xx) + \URth(\xx) = \Uelth(\xx) + \Umath(\xx).
	\label{eq:chEn}
\end{align}

\revChange{Accordingly, the left and right CPE components are derived
from the optical chirality and energy densities
as $\Uth_{\text{L}, \text{R}} = 1/2 \{ \Uth \pm c_0/(\omega n) \chith \}$.}
This picture enables the analysis of chiral near-fields with respect to their degree of local circular polarization.
It is known that for chiral matter, the study of enhancement of chirality should be compared to both electric and
magnetic energy \cite{choi2012}. 
Conventionally, design of nano-optical devices aims at increasing
the local enhancement factor $\hat{\chith}(\xx) = \chith(\xx) / |\chith_0|$
with chirality $\chith_0$ of the incident light \cite{schaeferling2012}.

However, by studying CPE, both achiral and chiral effects are taken into account:
In measurements investigating circular dichroism, the differential response of e.g.,\ a chiral molecule
to two distinct circular polarizations of the incident light is analyzed. In this manner,
the coupling to chiral fields is separated from achiral coupling.
The CPE complements this differential picture when studying near-fields of a single
polarization state. If purely enhancement of optical chirality was analyzed in this situation, a simultaneous
increase in the achiral response could be overseen.

Namely, an equal enhancement of optical chirality as well as electric and magnetic energy is generally possible, as shown in Equations 
 (\ref{eq:chDiff}) and (\ref{eq:chEn}).
Chiral responses could be enhanced as much as the achiral ones. By~employing the CPE analysis, an
enhancement of only chiral coupling is clearly separated from high achiral fields.
\revChange{Since optical chirality arises from the difference of the generally independent variables $\ULth$ and $\URth$,
these offer an additional degree of freedom for designing the desired field properties.}
Accordingly, the CPE is helpful for studying both the handedness and the enhancement of chiral near-fields
and their interaction with different enantiomers of chiral matter.

\section{Controllable Chiral Near-Fields in Helical Metamaterial}

We propose to study light matter interactions in chiral near-fields in the vicinity of metal helices. 
These fields can be conveniently tuned locally and spectrally in order to perform spectroscopic measurements. 
We demonstrate this tunability in a numerical study within an array of intertwined right-handed helices. 
The geometry of the helix array follows a previously investigated setup~\cite{kaschke2014} which was proposed 
as a broadband circular polarizer.
\revChange{Such devices showing chiral far-field responses are based on employing chiral
resonances instead of separate electric and magnetic resonances}~\revChangeT{\cite{pendry2004}}.
\revChange{Although} in this study we investigate how the near-field properties depend on illumination parameters,
\revChange{we show that analyzing chiral rather than achiral field quantities is also beneficial in the
vicinity of plasmonic structures}.

\begin{figure} [H]
	\centering
		\begin{tabular}{ccc}
			\includegraphics[width=0.2\textwidth]{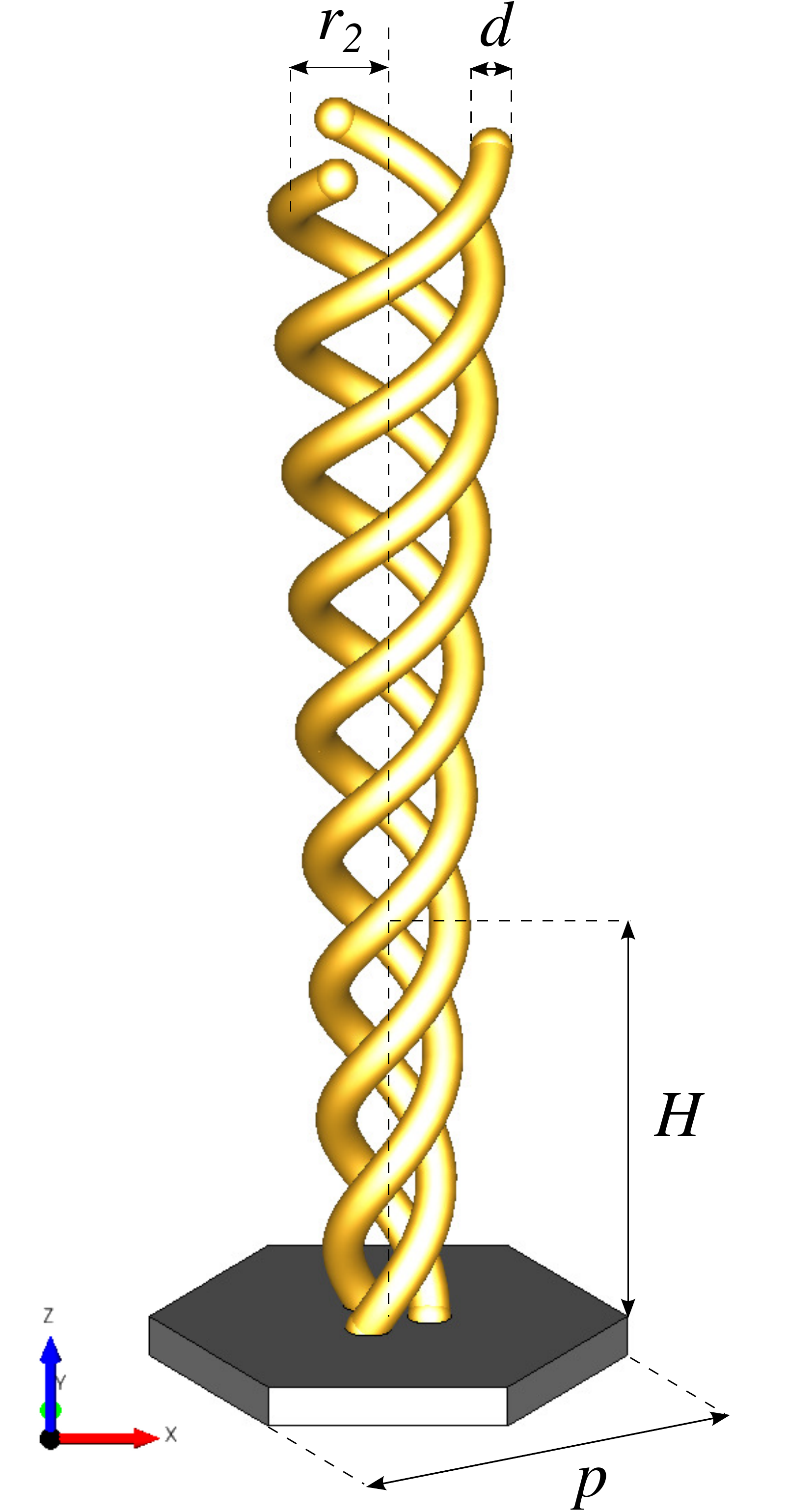}
			&
			\raisebox{2.5mm}{
				\includegraphics[width=0.2\textwidth]{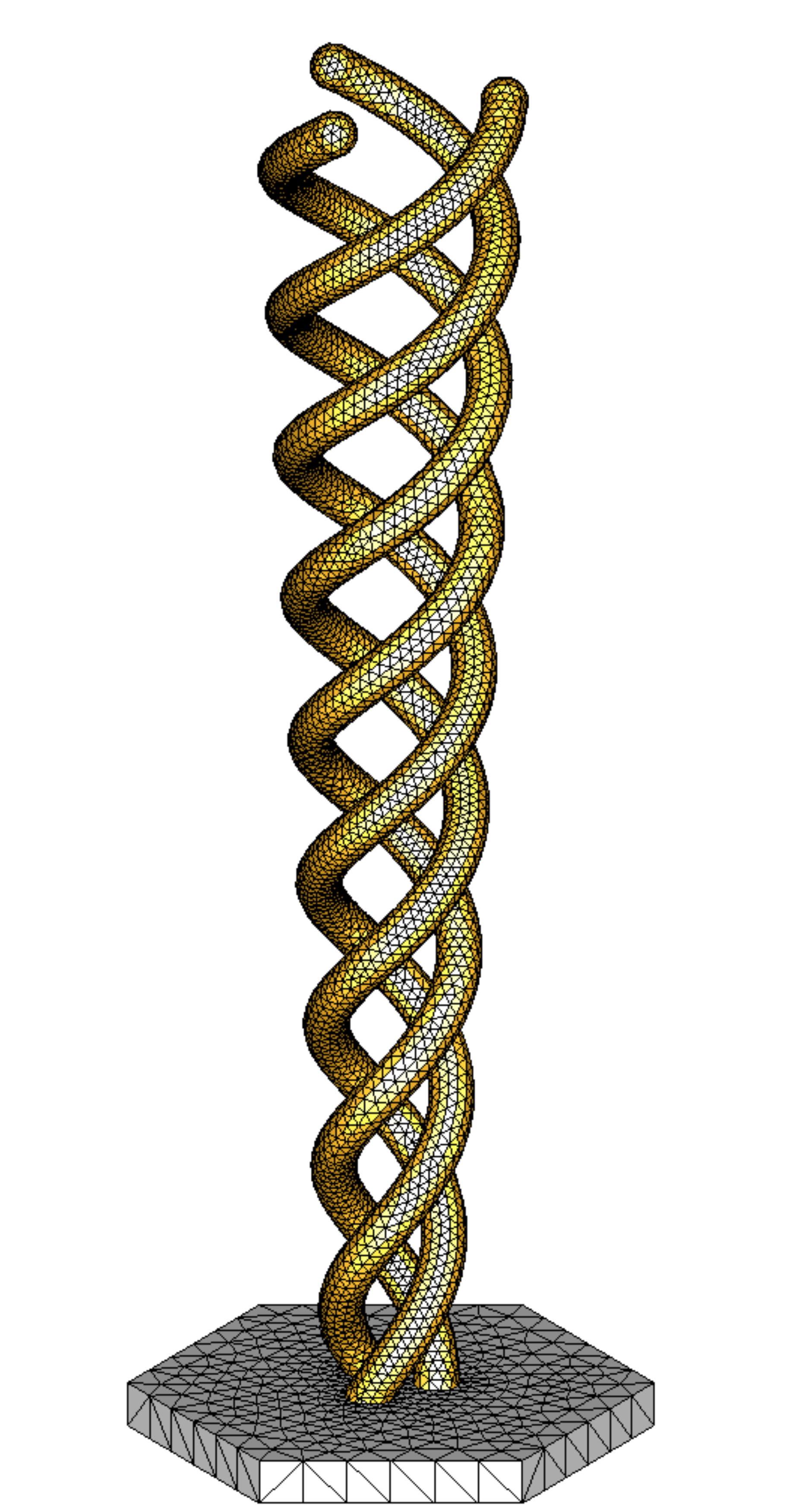}
			}
			&
			\raisebox{25mm}{
				\begin{tabular}{c}
					\includegraphics[width=0.38\textwidth]{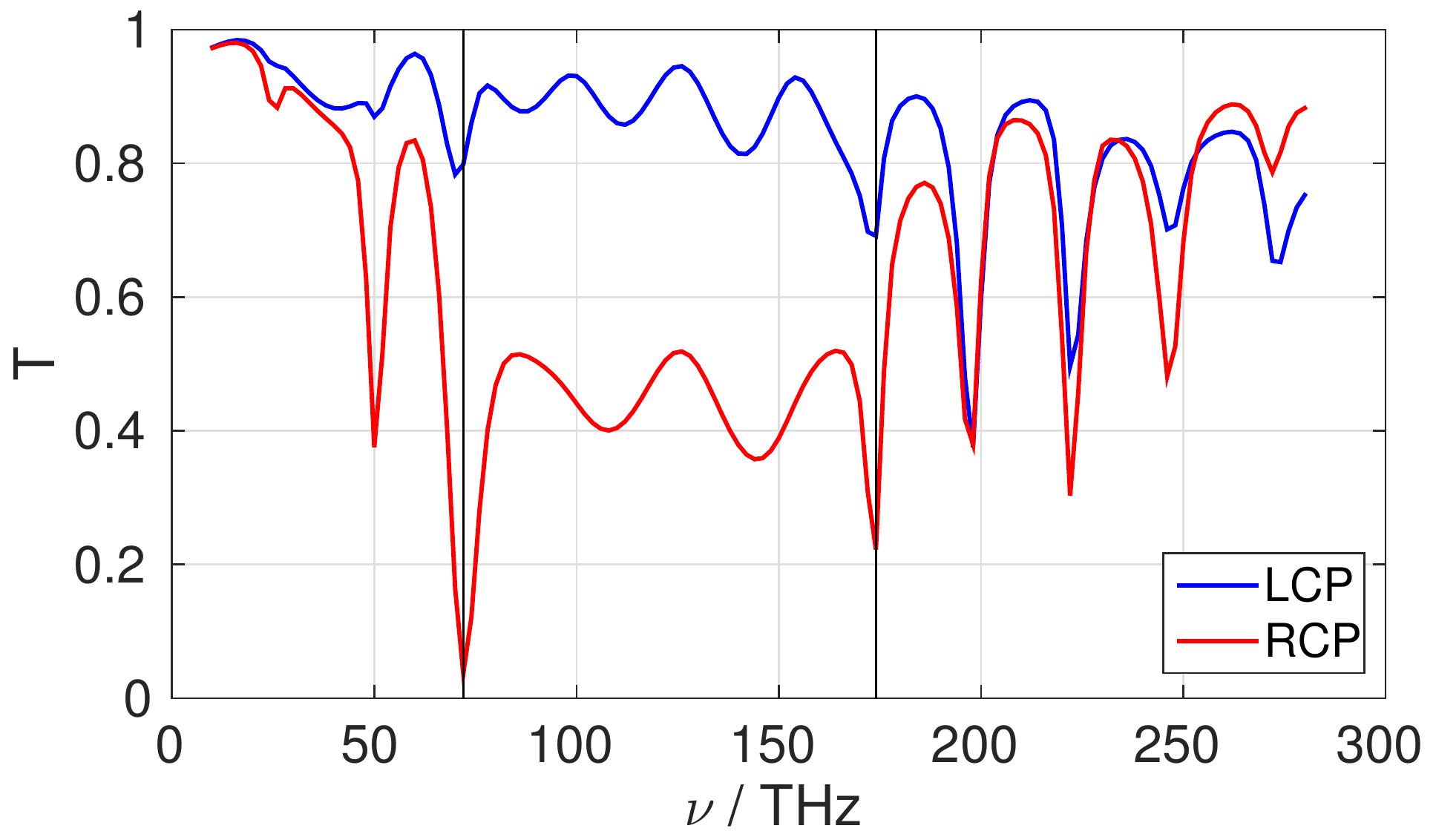}
					\\
					\includegraphics[width=0.38\textwidth]{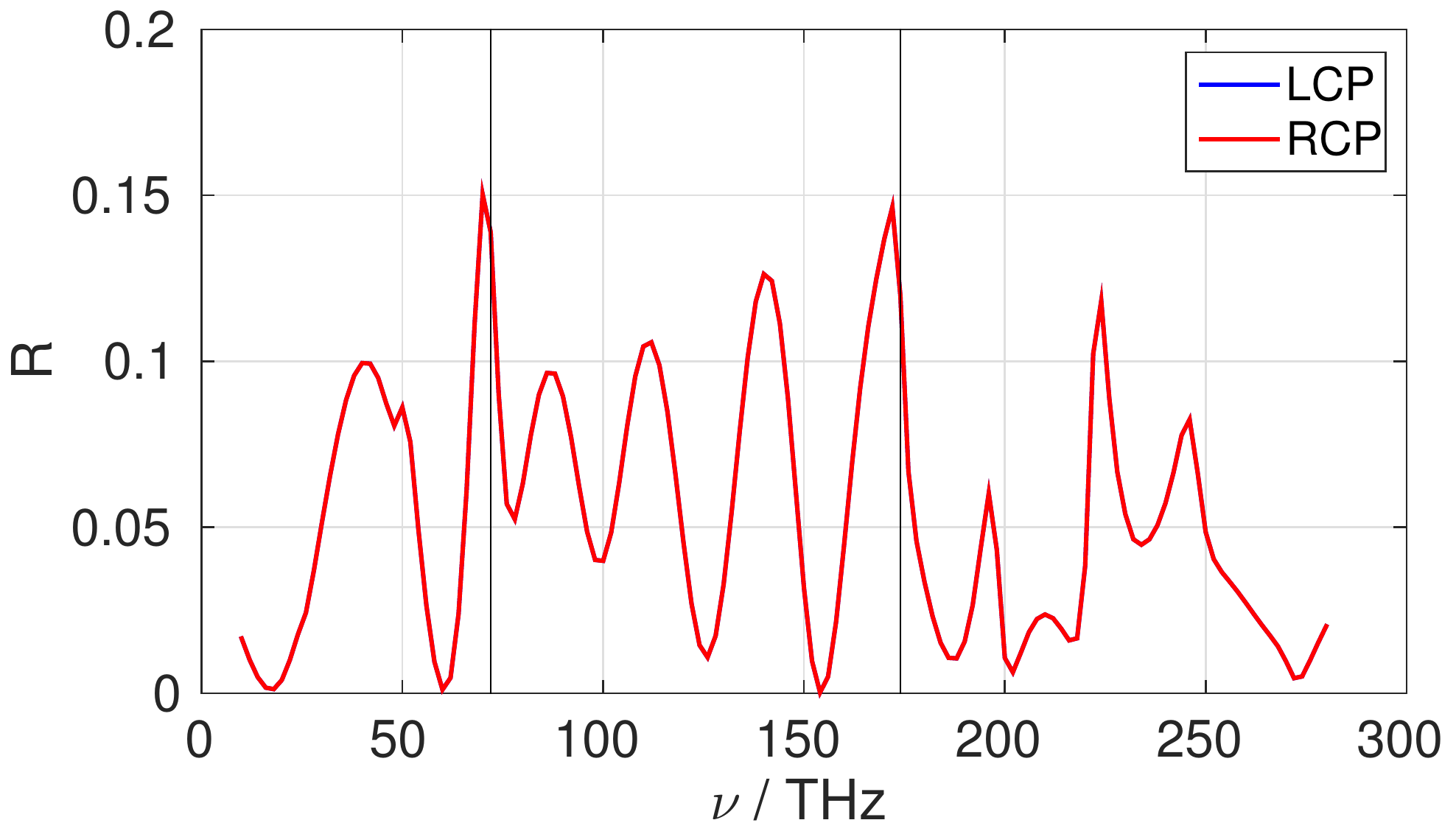}
				\end{tabular}
			}
			\\
			(\textbf{a}) & (\textbf{b}) & \revChange{(\textbf{c})}
		\end{tabular}

	\caption{
	\label{fig:setu}
	Tapered and intertwined gold 3-helices (\textbf{a}) and tetrahedral mesh (\textbf{b}) for finite-element method (FEM).
	Transmission $T$ (\textbf{c}, \textbf{top}) and reflection $R$ (\textbf{c}, \textbf{bottom}) spectra for normally incident right circularly polarized plane wave (RCP) 
	 (\textbf{red}) and left circularly polarized plane wave (LCP) (\textbf{blue}).
	Vertical black lines denote the band of suppressed $T$ for RCP
	.
	}
\end{figure}

Figure\ \ref{fig:setu}a shows a schematic view of the hexagonal unit cell of the helix array (pitch $p = 1 ~\upmu\text{m}$): 
Three intertwined, tapered metal helices are placed on top of a glass substrate and surrounded by air. 
The refractive index $n$ of the metal (gold) is given by a free-electron model 
with plasma frequency $\omega_\text{PL} = 1.37 \times 10^{16}$ rad/s
and collision frequency $\omega_\text{col} = 1.2 \times 10^{14}$ rad/s.

The helices' wire diameter is $d = 100~\text{nm}$, 
their axial pitch $H = 1~\upmu\text{m}$ and 
bottom and top radius $r_1 = 100~\text{nm}, r_2 = 250~\text{nm}$. 
In the simulations, the structure is illuminated from above 
with right/left circularly polarized plane waves (RCP/LCP) in the infrared spectral range, 
at inclination angle  $\theta$ with respect to the surface normal (helix axis) and rotation angle 
$\phi$ with respect to the $\Gamma$-$K$ direction of the array. Throughout this study, we set $\phi = 0$.

The structure is discretized with a tetrahedral mesh shown in Figure \ref{fig:setu}b.
Near-field distributions are computed using the finite-element method (FEM) 
implemented in the software package \textit{JCMsuite}~\cite{kaschke2014}.
Post-processes are used to determine electromagnetic field energies, chirality densities, transmission, reflection
and other quantities.
We have checked that the numerical relative error of the presented results is below one percent
with respect to conservation of energy as well as conservation of optical chirality.

Figure\ \ref{fig:setu}c shows reflection and transmission spectra for $\theta=0$. 
In accordance with symmetry considerations \cite{menzel2010,kaschke2012}, the reflection for RCP and LCP is equal, 
leaving absorption as the major mechanism for filtering polarization. 
The structure is resonantly excited by RCP,
yielding a band of suppressed transmission from approximately $72$ to $174$ THz. 
Here, mean transmission coefficients are smaller than $0.45$ for RCP and larger than $0.9$ for LCP.

\revChange{The tapering of the device results in a significantly increased operation bandwidth of the circular
polarizer} \revChangeT{\cite{kaschke2012}}\revChange{. For the near-field analysis employed here, the angle of the tapering
gives an~additional degree of freedom for optimizing the position and magnitude of the observed enhancements.
Especially for the tunability with respect to the obliqueness of the illumination described in Section {\ref{sec:angle}},
the~tapering yields a constant volume of increased fields for different angles of incidence.}

\subsection{Frequency-Dependence of Local Enhancement}

\begin{figure} [H]
\centering
		\includegraphics[width=.98\textwidth]{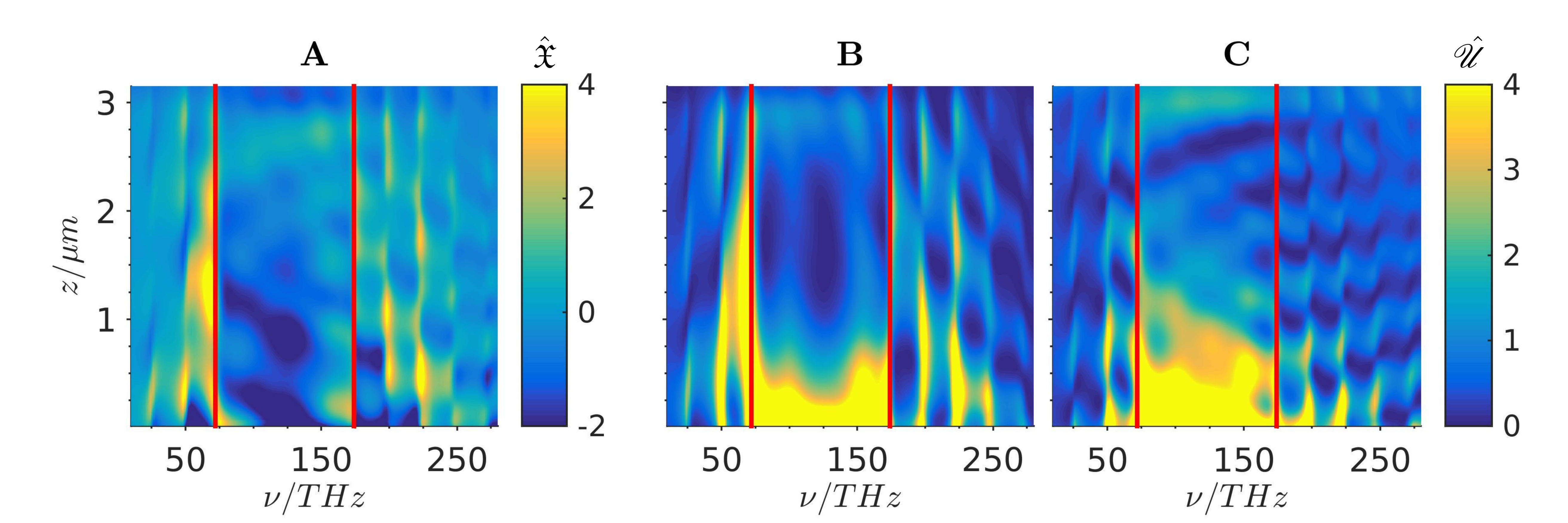}
	\caption{
	\label{fig:freq}
	Spectral near-field response of a 3-helix array illuminated with normally incident RCP. 
	Enhancements along the central $z$-axis are shown in pseudo-color representation
	as functions of frequency $\nu$. Vertical red lines denote the band gap.
	(\textbf{A}): chirality enhancement $\hat{\chith}(z,\nu)$.
	(\textbf{B},\textbf{C}):~enhancement of left [right] circularly polarized energy $\hat{\ULth}(z,\nu)$ 
	[$\hat{\URth}(z,\nu)$].
	\revChange{Note the predominantly positive chirality at the left red line.}
	}
	
\end{figure}

First, we analyze enhancement $\hat{\Uth_i}(z) = \Uth_i(z) / \Uth_{i,0}$
of the different energy quantities $\Uth_i$, given in Equation~\eqref{eq:chEn},
compared to their free space value $\Uth_{i,0}$ as well as chirality enhancement $\hat{\chith}(z)$
on the helix axis (in~$z$-direction). The helical metamaterial is illuminated with normally incident
and right circularly polarized plane waves. The frequency $\nu$ is varied from 10 to 280 THz.
As shown in Figure\ \ref{fig:freq}, several features arise in the near-fields.

The highest magnitudes of optical chirality are located within $1.5~\upmu$m above the substrate
in the center of the unit cell (Figure\ \ref{fig:freq}A).
The band of suppressed transmission is depicted by vertical red lines. Within this band,
two distinct and broadband enhancements are visible. Here, the incident optical chirality $\hat{\chith}_\text{inc} = -1$
is roughly doubled. The position of these low enhancements are spatially stable up to around 200 nm for variations
of the frequency by 50 THz.
This proposes an experimental setup wherein chiral enantiomers are placed in the center of the device coupling to locally enhanced chirality.
Tuning the incident frequency amplifies signals of spectroscopy or enables spatial analysis of macro-molecules. 

Stronger enhancements of optical chirality are observed outside the band of suppressed transmission.
Above the band, field enhancements follow the oscillatory behavior of the transmission spectrum, whereas
below the band, significant local chirality enhancement is observed by more than a~factor of four at $\nu = 72$ THz.
Note that the optical chirality changes sign compared to the incident RCP outside the band.

Analysis of the enhancement of the right and left CPE components is displayed in Figure\ \ref{fig:freq}B,C. It is
much more prominent for resonant frequencies. An increase by more than four times is located less than 500 nm above the substrate
within the full band. Above and below the band, CPE shows similar oscillatory behavior as optical chirality.
The interplay of right and left CPE is the reason for the opposite handedness of the incident circular polarization outside the band.

\begin{figure} [H]
\centering
		\raisebox{7.5mm}{
			\includegraphics[width=.11\textwidth]{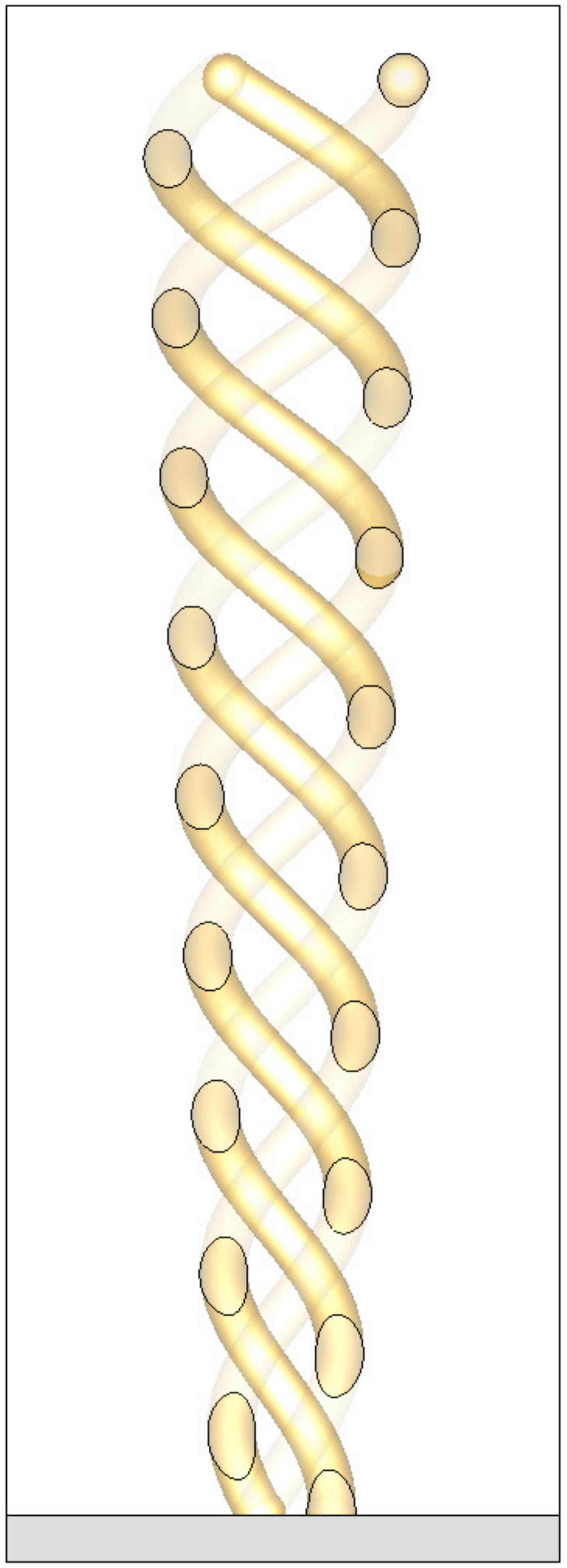}
		}
		\includegraphics[width=.82\textwidth]{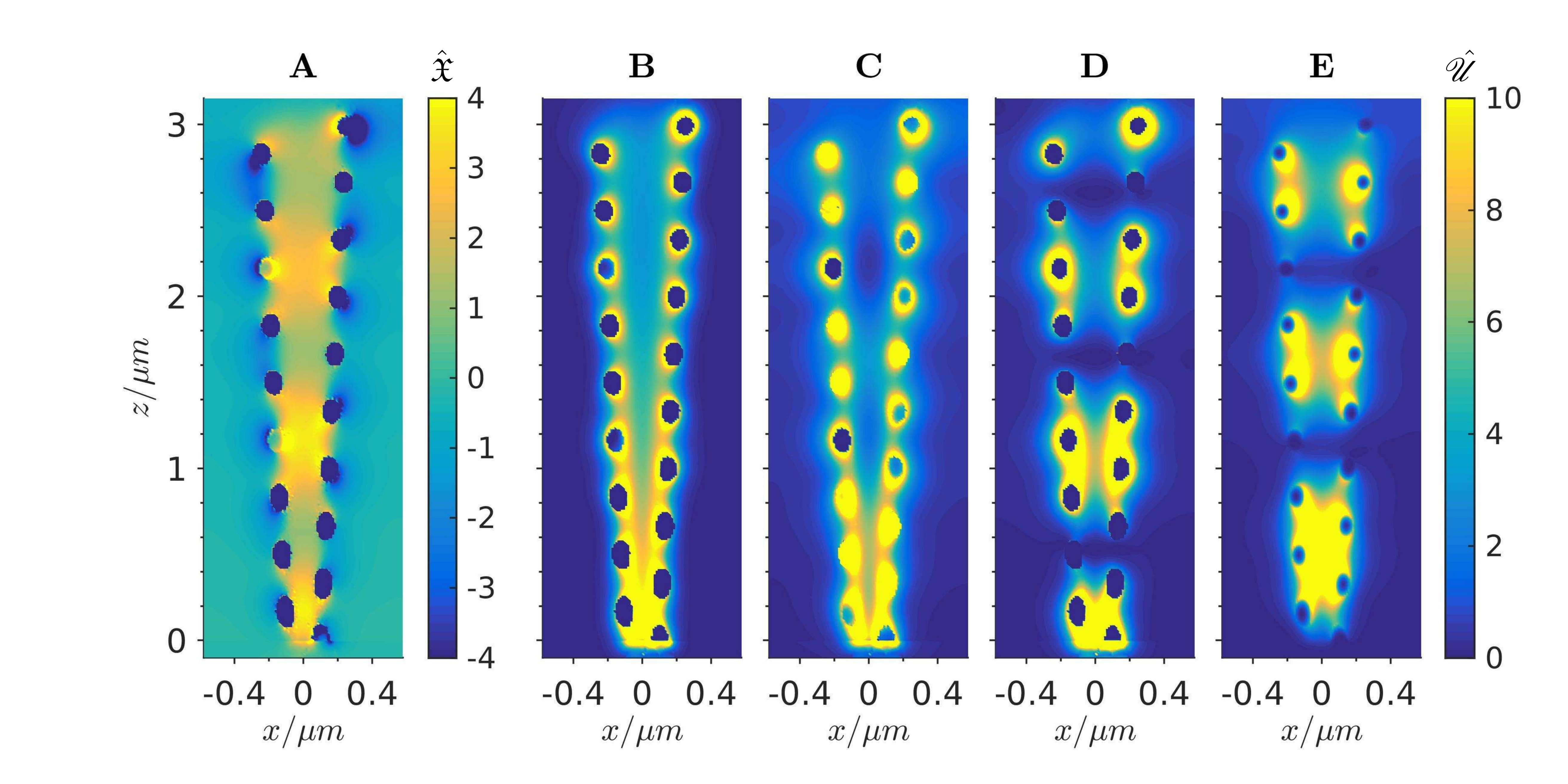}
	\caption{
	\label{fig:freq_E_chi}
	Near-fields of a 3-helix array at $\nu = 72$ THz illuminated with normally incident RCP.
	Enhancements compared to free space values in a $xz$-cross section (left) through
	the center of the unit cell are shown in pseudo-color representation.
	(\textbf{A}): chirality enhancement $\hat{\chith}(x,z)$;
	(\textbf{B},\textbf{C}): enhancement of left [right] circularly polarized energy $\hat{\ULth}(x,z)$
	[$\hat{\URth}(x,z)$];
	(\textbf{D},\textbf{E}): enhancement of electric (magnetic) field energy $\hat{\Uelth}(x,z)$
	[$\hat{\Umath}(x,z)$].
	}

\end{figure}

As an example, we show in Figure\ \ref{fig:freq_E_chi} the local optical chirality (A) as well as right/left (B/C) and electric/magnetic (D/E)
field energy enhancements for $\nu = 72$ THz. At this frequency, $\hat{\chith}(z,\nu)$ is maximal.
The electric and magnetic field energy densities
are inversely proportional as visible in Figure\ \ref{fig:freq_E_chi}D,E
\revChange{, i.e., when $\Uelth$ is maximum, $\Umath$ is minimum}.
The pronounced local chirality enhancement (A) is
due to a high left CPE $\ULth$ (B) while maintaining small $\URth$ (C).

In the picture of circularly polarized energy, these chiral near-fields are clearly understood since chirality is proportional to the difference
of left and right CPE (\ref{eq:chDiff}). However, they arise from the complex interplay of electric and magnetic field components.
Their local phase relations are not directly accessible in the analysis of electric and magnetic energy densities. Nevertheless, these are the underlying reasons
for the observed local chirality enhancement. From the point of view of chirality conservation, the incident chirality is locally
converted \cite{gutsche2016} to left-handed CPE, resulting in the observed strong chiral near-fields.

\revChange{It has been shown that the generation of the observed chiral near-fields can be explained by the current flows
induced in helix wire} \revChangeT{\cite{demetriadou2012}.}
\revChange{The currents are circulating around the core of the helix, yielding
a magnetic polarization. Additionally, some parts of the current flow parallel to the helix axis and induce an electric polarization.
If these two contributions show the correct phase relation, the~optical chirality density and accordingly the difference of the CPE parts is enhanced.}

\subsection{Spatial Control of Chirality Enhancement by Tuning the Incidence Angle}
\label{sec:angle}

Second, we study near-fields of the helical metamaterial when varying the angle of
incidence $\theta$ at a fixed frequency $\nu = 128$ THz.
The illuminating plane wave is left circularly polarized.
Although the structure is close to transparent for normally incident LCP (Figure\ \ref{fig:setu}c),
we observe high optical chirality and accurate control of the enhancement position.

\begin{figure} [H]
\centering
		\includegraphics[width=.98\textwidth]{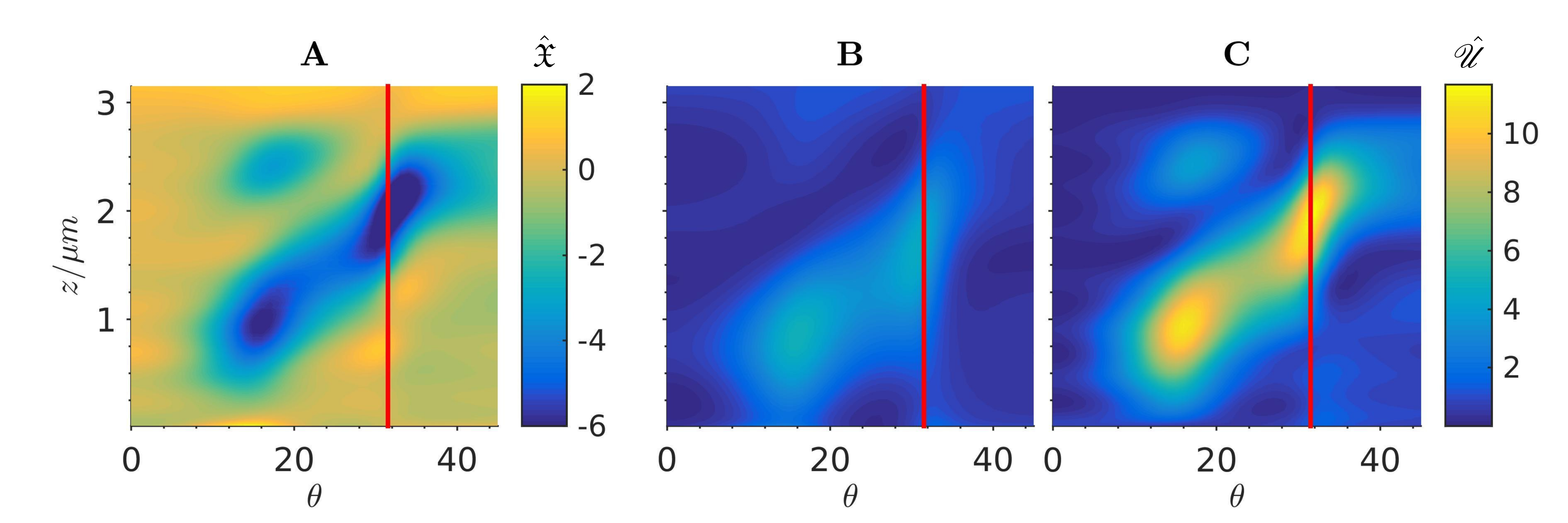}
	\vspace{-2ex}
	\caption{
	\label{fig:angle}
	Near-field response of a 3-helix array illuminated with LCP at $\nu = 128$ THz.
	Enhancements along the central $z$-axis are shown in pseudo-color representation
	as functions of incidence angle $\theta$. The vertical red line denotes the angle of incidence
	$\theta = 31.5^\circ$ with maximal $|\hat{\chith}| > 6$.
	(\textbf{A}): chirality enhancement $\hat{\chith}(z,\theta)$;
	(\textbf{B},\textbf{C}): enhancement of left [right] circularly polarized energy $\hat{\ULth}(z,\theta)$ 
	[$\hat{\URth}(z,\theta)$].
	\revChange{Note the predominantly negative chirality when tuning the angle.}
	}
\end{figure}

In Figure \ref{fig:angle}A, the chirality enhancement on the symmetry axis of the device for $\theta \in [0^\circ,~ 45^\circ]$ is shown.
The maximal enhancement is tunable between 1 and 2~$\upmu$m above the substrate. It is nearly proportional to the incidence angle.
In the range of $12^\circ$--$36^\circ$ the magnitude of the local chirality is increased more than four times with a maximal value of more than 
six depicted by a red vertical line.
This allows for easy experimental control of the position of high optical chirality at constant frequency and enables
the spatial analysis of chiral molecules.

\begin{figure} [H]
\centering
		\includegraphics[width=.98\textwidth]{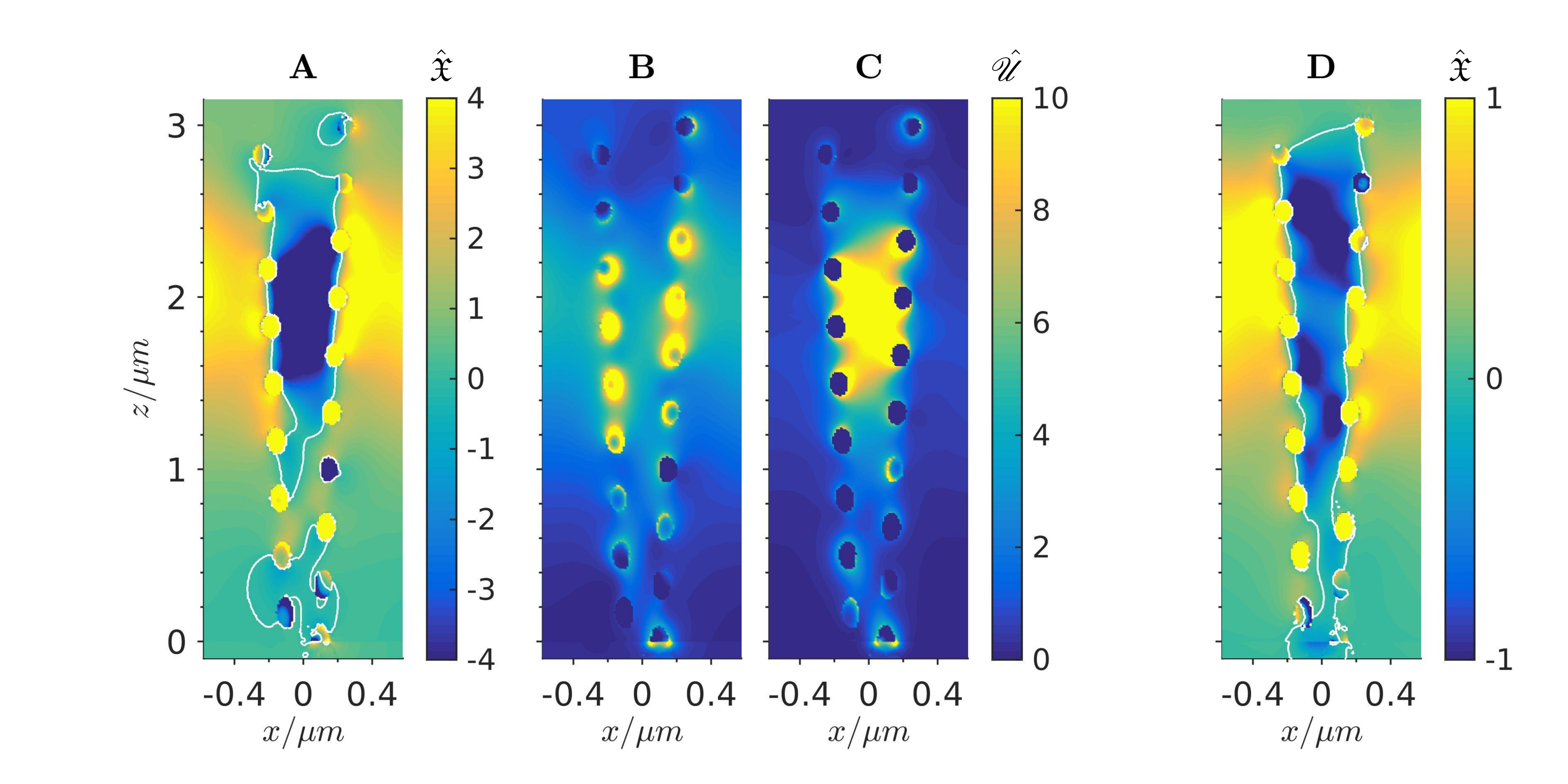}
	\vspace{-2ex}
	\caption{
	\label{fig:angle_scat_reso}
	Comparison of near-field response for scattering (\textbf{A}--\textbf{C}) with LCP at $\nu = 128~\text{THz}$ and $\theta = 31.5^\circ$
	and resonance (\textbf{D}) computation with corresponding
	Bloch vector.
	Enhancements in a $xz$-cross section through
	the center of the unit cell are shown in pseudo-color representation.
	White lines depict zero chirality.
	(\textbf{A},\textbf{D}): chirality enhancement $\hat{\chith}(x,z)$ from scattering [resonance] computation.
	(\textbf{B},\textbf{C}): enhancement of left [right] circularly polarized energy $\hat{\ULth}(x,z)$
	[$\hat{\URth}(x,z)$] from scattering simulation.
	}
\end{figure}

The strong chiral near-fields are due to an increased right CPE (Figure \ref{fig:angle}C)
and a relatively low variation of the left-handed energy part (B).
Analogous to the frequency-study in the previous section, the sign of the incident chirality is reversed.
In Figure\ \ref{fig:angle_scat_reso}A--C,
we show the spatial distribution of chirality and CPE enhancement for the angle with maximal chirality enhancement $\theta_\text{max} = 31.5^\circ$.
Here, the electric energy is much less increased whereas the magnetic part is strongly
enhanced (not shown).

Again, the concept of circularly polarized energy is shown to be suitable for chiral phenomena:
Although the electric dipole moment of most molecules is much stronger,
the effect of magnetic coupling cannot be neglected for chiral phenomena \cite{choi2012}.
On the other hand, studying CPE as given in Equation~\eqref{eq:chEn} is directly related to
the interaction of light and chiral matter. 
This is underpinned by the close connection 
of chirality and the general polarization ellipse of local fields \cite{bauer2015}.

Finally, we investigate the unexpected behavior of reversing the sign of the incident optical chirality.
As described in the previous section, this effect suggests a conversion mechanism of the metamaterial:
Locally, chirality or the circular polarization of the incident light is converted
into the opposite chirality. This occurs for resonant incident light showing low transmission.

We perform an analysis of accessible eigenmodes at the angle with maximal chirality enhancement $\theta_\text{max}$.
This is done by solving the resonance problem of Maxwell's equations with Bloch conditions corresponding to this specific angle.
Subsequently, the most suitable mode with local circular polarization and eigenfrequency similar to the externally excited near-field is chosen.

Generally, the coupling to all eigenmodes which overlap with the external field has to be taken into account
for a full description of the device. Additional interference of the incident field and different modes
might occur.
However, in order to show that the tuning capabilities, with respect to the angle
of incidence, are based on modes accessible only under oblique illumination, we analyze the eigenfield
which is mostly excited.

In Figure\ \ref{fig:angle_scat_reso}A,D,
we compare near-field chirality of the illumination with LCP
with the corresponding eigenmode.
Since its eigenfrequency and the frequency of the incident plane wave is equal, the incident light couples to this mode,
although its local chirality is of opposite handedness.
Accordingly, the observed chirality conversion mechanism occurs. We expect a similar behavior in the case of the frequency-dependent chirality enhancement
in the previous section.

\section{Conclusions}

In summary, metamaterials composed of intertwined and tapered 3-helices
enable the control of the chiral optical near-fields both spatially and in magnitude.
Variation of wavelength or the angle of incident CPL controls
the vertical position in the range of micrometers of local chirality. Enhancement factors of up to six are obtained
on the central axis of this device.
The numerical analysis, with the help of CPE, simplifies the complex interplay of electromagnetic fields forming local circular
polarization and gives design guidelines
for chiral light-matter interactions.

\revChange{We expect the two independent CPE components to offer more degrees of freedom for optimization of comparable devices
than solely the quantity of optical chirality and have shown that the interplay of extrinsic chiral parameters, such as angle
of incidence and intrinsic geometric chirality of e.g., tapered 3-helices, offer the ability to tune chiral near-fields.}

\vspace{6pt} 

\acknowledgments{
We acknowledge support from the BMBF through project 13N13164,
the Einstein Foundation Berlin through project ECMath-SE6,
the DFG through SFB787-TP4
and the Freie Universit\"at Berlin.
}

\authorcontributions{
All authors conceived the concept of the study and discussed the results.
R.M. performed the simulations.
P.G. wrote the paper with input from S.B. and R.M.
}

\conflictofinterests{The authors declare no conflict of interest.}

\abbreviations{The following abbreviations are used in this manuscript:\\

\noindent 
\begin{tabular}{@{}ll}
CPL & Circularly Polarized Light\\
CPE & Circularly Polarized Energy\\
RCP & Right Circularly Polarized Plane Wave\\
LCP & Left Circularly Polarized Plane Wave\\
FEM & Finite-Element Method
\end{tabular}}

\bibliographystyle{mdpi}

\newpage
\renewcommand\bibname{References}


\end{document}